\documentclass[twocolumn,aps,reprint,shownopacs,superscriptaddress]{revtex4}
\usepackage{graphicx}
\usepackage{amssymb}
\usepackage{amsmath}
\usepackage{amsthm}
\usepackage{array}
\usepackage[colorlinks=true,linkcolor=blue,urlcolor=blue,citecolor=blue,pdfauthor={ },pdftitle={},pdfsubject={ },pdfkeywords={ }]{hyperref}

\graphicspath{{Figures/}}

\begin{document}

\title{Optimizing adiabatic quantum pathways via a learning algorithm}
\author{Xiaodong Yang}
\affiliation{CAS Key Laboratory of Microscale Magnetic Resonance and Department of Modern Physics, University of Science and Technology of China, Hefei, Anhui 230026, China}

\author{Ran Liu}
\affiliation{CAS Key Laboratory of Microscale Magnetic Resonance and Department of Modern Physics, University of Science and Technology of China, Hefei, Anhui 230026, China}

\author{Jun Li}
\email{lij3@sustech.edu.cn}
\affiliation{Shenzhen Institute for Quantum Science and Engineering and Department of Physics, Southern University of Science and Technology, Shenzhen 518055, China}

\author{Xinhua Peng}
\email{xhpeng@ustc.edu.cn}
\affiliation{CAS Key Laboratory of Microscale Magnetic Resonance and Department of Modern Physics, University of Science and Technology of China, Hefei, Anhui 230026, China}
\affiliation{Hefei National Laboratory for Physical Sciences at the Microscale, University of Science and Technology of China, Hefei 230026, China}
\affiliation{Synergetic Innovation Centre of Quantum Information $\&$ Quantum Physics, University of Science and Technology of China, Hefei, Anhui 230026, China}

\date{\today}

\begin{abstract}
Designing proper time-dependent control fields for slowly varying the system to the ground state that encodes the problem solution is crucial for adiabatic quantum computation. However, inevitable perturbations in real applications demand us to accelerate the evolution so that the adiabatic errors can be prevented from accumulation. Here, by treating this trade-off task as a multiobjective optimization problem, we propose a gradient-free learning algorithm with pulse smoothing technique to search optimal adiabatic quantum pathways and apply it to the Landau-Zener Hamiltonian and Grover search Hamiltonian. Numerical comparisons with a linear schedule, local adiabatic theorem induced schedule, and gradient-based algorithm searched schedule reveal that the proposed method can achieve significant performance improvements in terms of the adiabatic time and the instantaneous ground-state population maintenance. The proposed method can be used to solve more complex and real adiabatic quantum computation problems.
\end{abstract}
\maketitle

\renewcommand{\thesubsection}{\arabic{subsection}}

\section{Introduction}
Adiabatic quantum computation (AQC) \cite{AL18}, which is known to be polynomially equivalent \cite{AV08} to the standard circuit-based quantum computation, offers us an alternative way to solve many challenging optimization problems, such as the traveling salesman problem \cite{MR04} and satisfiability problem \cite{FG01}. It functions by designing a target Hamiltonian whose ground state encodes the solution of the optimization problem of interest, and slowly evolving the system to this target Hamiltonian from some simple initial Hamiltonian whose ground state can be easily prepared. According to the quantum adiabatic theorem \cite{FG00,FG01}, as long as the system evolves sufficiently slowly and the external uncertainties have only negligible effects on the system, the final state of the system will be the ground state of the target Hamiltonian, as expected. 

In actual applications, although AQC has inherent robustness to some sources of noise, such as dephasing and unitary control errors \cite{CF01,AA09}, its effectiveness can still be severely hampered by other inevitable perturbations. Consequently, many-error suppression and error-correction methods \cite{JF06,LD08,YS13,SY13} have been developed to handle this problem. However, recent study \cite{YS13} shows that these methods are not sufficiently fault tolerant, and they are rather resource consuming. A more practical and direct approach is to design a sufficient fast adiabatic evolution path, for the sake of reducing the accumulations of the adiabatic errors.

Shortcuts to adiabaticity \cite{GR19} are a representative approach to accelerate the transition to the target state, but it always needs complicated analytical derivations, detailed information of instantaneous adiabatic state of the system, or unfeasible additional terms \cite{TK19}. Furthermore, it inherently cannot maintain the instantaneous ground state during the evolution process, and thus is not proper for most of the AQC applications. Recent efforts have brought new opportunities to adiabatically accelerate the evolution by optimal control methods, including analytical quantum adiabatic brachistochrone (QAB)\cite{RK09}, numerical Lyapunov control \cite{WH12}, and gradient-based methods \cite{BG14,QG19}. However, QAB is only suitable for low-dimensional parametrizations and does not consider the population loss during the evolution process. Gradient-based methods greatly rely on initial trial controls, their derivatives require an abundant amount of resources to obtain, and they are more easily trapped into the local extremes for complex optimization problems \cite{PS06}.

Here, we formulate this task, i.e., decreasing the adiabatic time while minimizing the population loss from the instantaneous ground state, as a multiobjective optimization problem. We employ a simple but powerful differential evolution (DE) \cite{SP97,DSP11,DMS16} algorithm to explore the tradeoffs between these two objectives. Such gradient-free learning algorithms have drawn much attention in recent studies for their ability to produce high-quality controls and design better experiments \cite{RSHR92,BR10,RLR13,ZG15,ZG16,KMF16,FUZ17,YL19}. In this study, specifically, we consider cases where all the controls in the time-dependent Hamiltonian can vary freely, but with amplitude constraints, and the objective function to be maximized contains two weighted terms: the target state fidelity and the averaged system energy during the evolution. These multi objective optimization problems with constraints are very hard to solve analytically. Compared to a recent work using a gradient-based algorithm (called D-MORPH) and instantaneous ground-state tracking \cite{BG14} to solve this multi objective problem, our approach promises larger probability in finding global optimal solutions and is more practical to iteratively implement in real experiments. As illustrative applications, we perform numerical demonstrations on a Landau-Zener Hamiltonian \cite{BV12} and Grover search Hamiltonian \cite{GR97} using the proposed approach. Comparisons are also made to show the advantages of our approach over the above-mentioned gradient-based method. Further practical and complex applications of our method for AQC computation are also briefly discussed.

The remainder of the paper is organized as follows. We first introduce the AQC basics and formulate the pathway optimization problem in a general setting in Sec. \ref{Problem}. The learning algorithm for AQC is then described in Sec. \ref{DE}. Afterwards, we choose two representative pathway optimization problems and compare the numerical simulation results of our proposed method and some previously reported methods in the literature in Sec. \ref{Numeric}. Finally, in Sec. \ref{Conclusion}, some brief conclusions and discussions are presented.

\section{Backgrounds and problem setup } \label{Problem}
Consider an $n$-qubit quantum system which evolves under the following Schr\"odinger equation  ($\hbar=1$):
\begin{equation}
\frac{d|\psi(t)\rangle}{dt}=-i\mathcal H(t)|\psi(t)\rangle, \quad t\in[0,T],
\end{equation} 
where $\mathcal H(t)$ represents the time-dependent system Hamiltonian and the Hilbert space dimension is $N=2^n$. Thus, the quantum state $|\psi(t)\rangle$ can be transformed with $|\psi(t)\rangle=U(t)|\psi(0)\rangle$, where the evolution operator $U(t)$ satisfies
$dU(t)/{dt}=-i\mathcal H(t)U(t),U(0)=I.
$
The instantaneous eigenstates and eigenenergies of $\mathcal H(t)$ can then be defined by 
\begin{equation}
	\mathcal H(t)|\phi_m(t)\rangle=E_m(t)|\phi_m(t)\rangle 
\end{equation}
with $ m=0,1,...,N-1$ and $E_0(t) \leq E_1(t) \leq...\leq E_{N-1}(t)$. Here, we are mainly concerned with the energy gap between the ground state and the first-excited state, i.e., $g(t) = E_1(t)-E_0(t)$.

To perform AQC, the routine is to first prepare the system at the ground state $|\psi(0)\rangle=|\phi_0(0)\rangle$ of the initial Hamiltonian $\mathcal{H}_I=\mathcal{H}(0)$, which is assumed to be easily prepared. The system then evolves slowly under the constructed Hamiltonian, 
\begin{equation}
\mathcal H(t)=\mathcal H[\mathbf{u}(t)]= u_1(t)\mathcal H_I+u_2(t)\mathcal H_P, 
\end{equation}
where $\mathcal H_P=\mathcal H(T)$ represents the problem Hamiltonian, and $u_1(t),u_2(t)$ are control fields satisfying the boundary conditions $u_1(0)=u_2(T)=1,u_1(T)=u_2(0)=0$ and amplitude constraints $0\leq u_l(t)\leq 1,l=1,2$. As designed, the ground state $|\phi_0(T)\rangle$ of $\mathcal H_P$ encodes the solution to the computational problem. The quantum adiabatic theorem \cite{FG00,FG01} guarantees that as long as the evolution is sufficiently slow and the external perturbations can be ignored, the system's final state $|\psi(T)\rangle$ will be the target ground state $|\phi_0(T)\rangle$. To quantify their distance, we define the state fidelity $F_1=|\langle\phi_0(T)|\psi(T)\rangle|^2$. 

 The control schedules $\mathbf{u}(t)=(u_1(t),u_2(t))$ that dominate the above system evolution, which we called adiabatic quantum pathways, are very crucial for the reliable realization of AQC. Different methods have been developed to design or search such controls, as mentioned before. For the following comparisons with our proposed method, here we briefly review two conventional methods. The first one is to use linear interpolation control fields \cite{RC02} (marked as Linear), i.e., $u_2(s)=1-u_1(s)=s$, where we use the rescaled time $s=t/T$. Another one is based on the local adiabatic evolution theorem \cite{AL18,RC02} (RC for short); for the Grover search Hamiltonian, it is $u_2(s)=1-u_1(s)=1/2+\tan [(2s-1) \tan ^{-1} \sqrt{N-1}]/{2 \sqrt{N-1}}$, with $s=t/T$.

\section{Differential evolution algorithm for AQC}\label{DE}
To numerically optimize the control schedules using the differential evolution algorithm, we should first set a performance function to evaluate these controls. As mentioned previously, we use a multiobjective function as follows \cite{BG14}:
\begin{eqnarray}\label{obj}
	F=|\langle\phi_0(T)|\psi(T)\rangle|^2 
	-\frac{\alpha}{T} \int_0^T  \langle \psi(t)| \mathcal H(t)|\psi(t)\rangle dt, 
\end{eqnarray}
where $\alpha>0$ is a positive weight factor that determines the relative importance of the first term ($F_1$), which represents the main physical goal, and the second term (with minus, denoted as $F_2$), which is used to minimize the population loss from the instantaneous ground state during the evolution. Additionally, to quantify the instantaneous population loss, we define the instantaneous ground-state population $P_0(t)=|\langle\phi_0(t)|\psi(t)\rangle|^2 $.

\begin{figure*}
  \centering
\includegraphics[width=0.8\textwidth,height=0.48\textwidth]{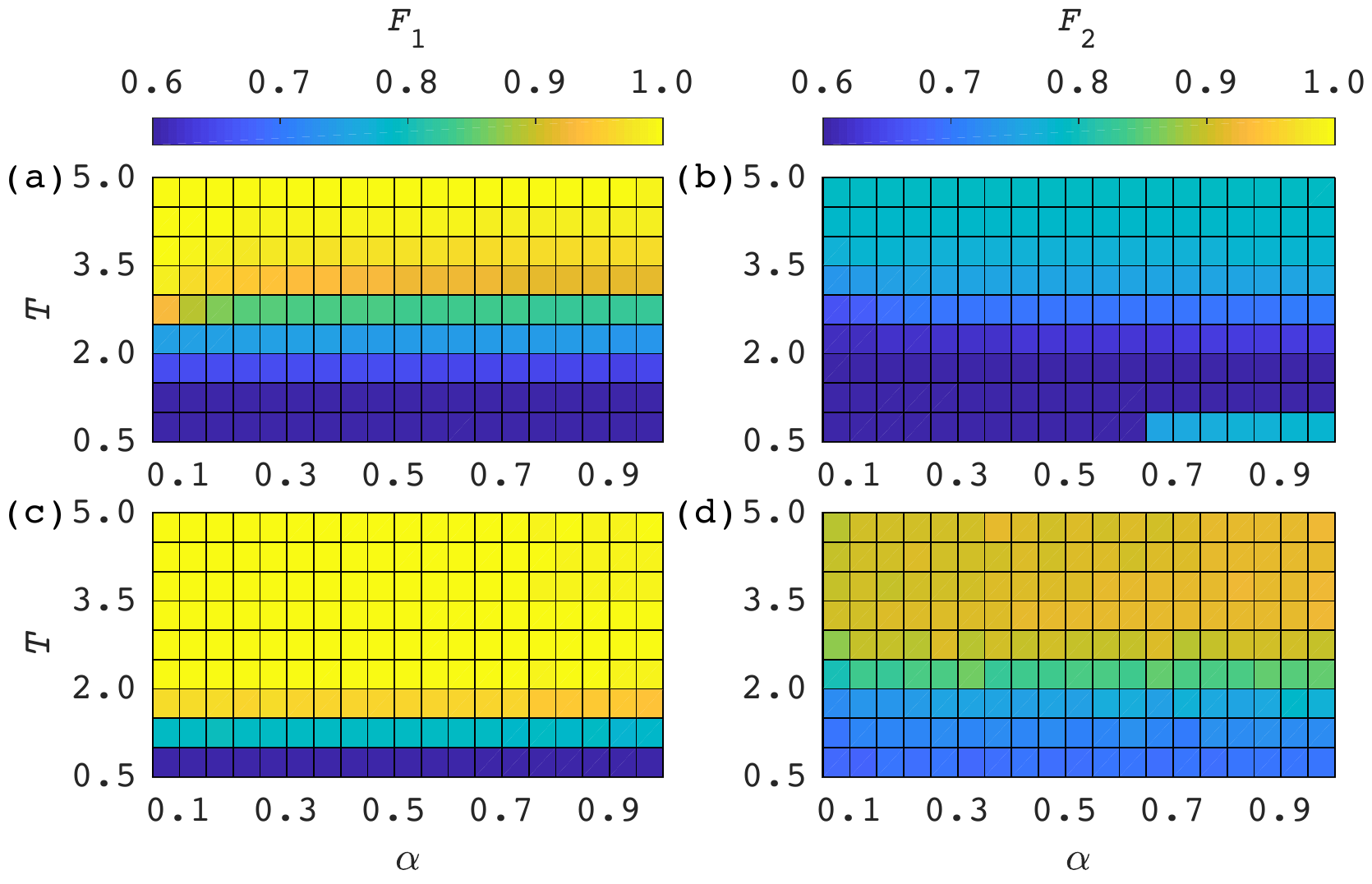}\\
  \caption{(Color online) Performance function values $F_1$ and $F_2$ vs different combinations of the adiabatic time $T$ and the weight factor $\alpha$ for a Landau-Zener Hamiltonian. (a), (b) The averaged results obtained by D-MORPH over five runs. The maximum iteration number was $G_{\max}=1000$. The step size $\lambda^G$ was initialized as $0.02$ and decreased by a factor $0.5$ if the calculated $F$ was worse than the previous one, but with the maximum trial times $100$. The control fields were all bounded in the range $[0,1]$ during the optimization. (c), (d) The averaged results produced by DE over five runs. The maximum iteration number was $G_{\max}=300$, and the initial guess was chosen as $u_1^g(s)=1-s,u_2^g(s)=s$. The algorithm parameters were $S=0.6,C=0.95,P=20,D=12,N_c=2$. Moreover, the controls were also constrained in the range $[0,1]$ during the searching process.}\label{Figure1}
  \end{figure*}

The optimal control schedules should not only maximize the above multi objective function, but also be smooth enough so that the real applications can realize predicted performance. To achieve this, we use the chopped random basis (CRAB) technique \cite{CCM11} to express the controls to be optimized in a set of truncated  Fourier basis,
\begin{eqnarray}
u_l(s)=u_l^{\text{g}}(s)\{ 1+{\sum_{k=1}^{N_c}[a_l^k \sin(\omega_l^k s)+b_l^k \cos(\omega_l^k s)]} \},
\end{eqnarray}
where we use the scaled time $s=t/T$, $l=1,2$ indicate the index of the two control fields, and $u_l^{g}(s)$ represents the initial controls guess. 
Thus, the optimization of the control schedules, $\mathbf{u}(s)=(u_1(s),u_2(s))$ is to search $6N_c$ optimal parameters $X=(a_1^k,b_1^k,\omega_1^k,a_2^k,b_2^k,\omega_2^k)$ ($k=1,2,\cdots,N_c$) that maximize the above performance function given by Eq. (\ref{obj}). In addition, to perform amplitude constraints on the control fields, we use the unity-based normalization, i.e., $u_l'(s)=(u_l(s)-{u_l}^{\max})/({u_l}^{\max}-{u_l}^{\min})$, where ${u_l}^{\max}$ and ${u_l}^{\min}$ represent the maximum amplitude and the minimum amplitude of $u_l(s):s\in[0,1]$, respectively.

Differential evolution algorithm \cite{SP97,DSP11,DMS16}, as a simple but competitive real-valued gradient-free optimization method, is applied here to optimize these parameters. It functions by simulating the natural evolution process through applying the steps of operator mutation, crossover, and selection in the population space, which is made up of a set of individuals. The detailed algorithm procedures are described as follows.

\textit{Step 1}. Set the algorithm parameters: scaling factor $S$, crossover rate $C$, chromosome length (the dimension of each individual) $D$, and population size $P$. Generate an initial population $Pop= \{ X_1^0,...,X_{P}^0\}$ randomly, with $X_i^0 = [X_{i1}^0,...,X_{iD}^0]$ being the $i$ th individual in the current population.

\textit{Step 2}. Update the iteration number $G=G+1$, and from $i=1$ to $P$, do the following steps:

\noindent
(1) {Mutation}.  Generate a donor vector $V_i^G=[V_{i1}^G,...,V_{iD}^G]$ through the differential mutation scheme of DE: ${V_i^G} = {X_{r_b^i}^{G-1}} + S \cdot ({X_{r_1^i}^{G-1}} - {X_{r_2^i}^{G-1}})+ S \cdot ({X_{r_3^i}^{G-1}} - {X_{r_4^i}}^{G-1})$, where $r_1^i, r_2^i, r_3^i, r_4^i$ are randomly chosen, mutually exclusive integers in the range $[0, P]$, and $r_b^i$ is the index of the best individual in the current population.

\noindent
(2) {Crossover}. Generate a trial vector $U_i^G=[U_{i1}^G,...,U_{iD}^G]$ by binomial crossover strategy: if $rand_{i,j}[0,1] \le C$ or $j=j_{rand}$, let $U_{ij}^G=V_{ij}^G$, where $j_{rand} \in [1,2,...,D]$ is a randomly chosen index. Otherwise, let $U_{ij}^G=X_{ij}^{G-1}$.

\noindent
(3) {Selection}. Evaluate the former individual $X_i^{G-1}$ and the trail vector $U_i^G$: if $f(U_i^G)\le f(X_i^{G-1})$, let $X_i^{G}=U_i^G$. Otherwise, keep $X_i^{G}=X_i^{G-1}$ unchanged.

 \textit{Step 3}. Check the stopping criterion, and if not satisfied, go to \textit{Step 2}.

\begin{figure*}
  \centering
\includegraphics[width=0.85\textwidth,height=0.5\textwidth]{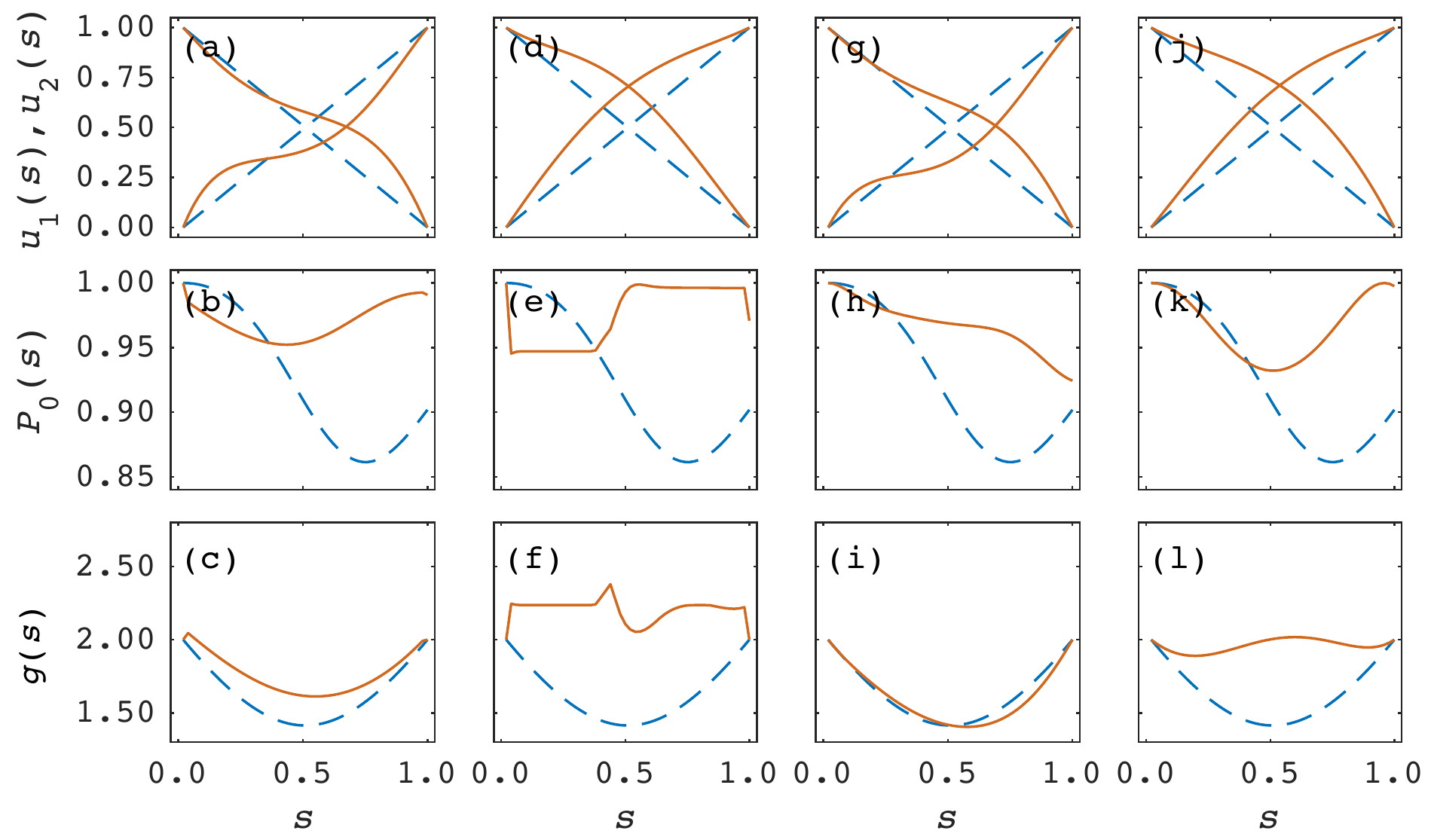}\\
  \caption{(Color online) Optimization results for the Landau-Zener Hamiltonian obtained by the Linear, D-MORPH, and DE methods. The controls fields $u_1(s)$ and $u_2(s)$, the instantaneous ground-state population $P_0(s)$, and the energy gap $g(s)$ are shown vs the scaled time $s$ optimized by (a)--(c) D-MORPH (solid red line) and Linear (dashed blue line) when $T=3,\alpha=0.1$; (d)--(f) DE (solid red line) and Linear (dashed blue line) when $T=3,\alpha=0.1$; (g)--(i) D-MORPH (solid red line) and Linear (dashed blue line) when $T=3,\alpha=0.5$; and (j)--(l) DE (solid red line) and Linear (dashed blue line) when $T=3,\alpha=0.5$.}\label{Figure2}
  \end{figure*}
We will also compare our method with the recent presented gradient-based D-MORPH \cite{BG14} method. In D-MORPH, the new controls can be refreshed iteratively by 
\begin{eqnarray}
u_l^{G+1}(t)=u_l^{G}(t)+\lambda^{G}{\partial F}/{\partial u_l^{G}(t)}
\end{eqnarray}
until the stopping criterion is met, where $\lambda^{G}$ is some appropriate step size, and ${\partial F}/{\partial u_l^G(t)}$ is the functional derivative of the objective with respect to each control field.
 
\section{Applications}\label{Numeric}
To show the advantages of our proposed method, we chose two representative examples, i.e., a Landau-Zener-type Hamiltonian \cite{BV12} and Grover-search-algorithm-type Hamiltonian \cite{GR97}, to demonstrate the numerical simulations.
\subsection{Landau-Zener Hamiltonian}
As a simple but nontrivial start-up, we explored the adiabatic quantum pathways of Landau-Zener Hamiltonian $\mathcal H_I=\sigma_z,\mathcal H_P=\sigma_x$, where $\sigma_x$ and $\sigma_z$ are Pauli matrices. Adiabatic time $T$ is crucial for the realization of AQC, it should be set carefully so that the system can evolve sufficiently slowly but without accumulating too many adiabatic errors. Moreover, as we use a multi-objective function to adjust the control schedules for optimal adiabatic quantum pathways, the weight factor $\alpha$ is very important for the success of the optimization. 

Therefore, we first studied the performance function values $F_1$ and $F_2$ with respect to different combinations of $T$ and $\alpha$ for the D-MORPH and DE methods, as shown in Fig. \ref{Figure1}. Here, a sufficiently large iteration number was set for both of the methods so that the best performance function values could be reached in each case with the settled $T$ and $\alpha$. From the comparison between Figs. \ref{Figure1}(a) and \ref{Figure1}(c), a direct and general conclusion is that DE performs better than D-MORPH for realizing the main physical goal, i.e., DE results in a final state closer to the ground state of the problem Hamiltonian. In more detail, we find that when $T$ is greater than $3$, D-MORPH has a comparable performance with DE for most of the weight factor $\alpha$. However, when $T$ is smaller than $3$, DE can still achieve a very high state fidelity $F_1$ for most of the cases but D-MORPH fails. Additionally, if we focus on the issue of how the weight factor $\alpha$ affects $F_1$, we can see that the performance of D-MORPH is much more sensitive to the choice of $\alpha$ than that for DE, and smaller $\alpha$ is more likely to achieve a better performance for D-MORPH. 
Besides maximizing the main goal $F_1$, we also care about minimizing the population loss during the optimization process. The comparisons in Figs. \ref{Figure1}(b) and \ref{Figure1}(d) reveal that DE also performs better than D-MORPH for optimizing $F_2$, especially for large adiabatic times and small weight factors. These results in Fig. \ref{Figure1} indicate that when searching optimal adiabatic quantum pathways for the Landau-Zener Hamiltonian, DE has great advantages over D-MORPH for a wide range of parameters $T$ and $\alpha$. To make this more concrete, we quantitatively compare these two methods and show some typical results in Table \ref{table1}.  

\renewcommand\arraystretch{1.1}
\begin{table}
\begin{tabular}{|p{0.8cm}<{\centering}|p{1.1cm}<{\centering}|p{1.1cm}<{\centering}|p{1.1cm}<{\centering}|p{1.1cm}<{\centering}|p{1.1cm}<{\centering}|p{1.1cm}<{\centering}|}
\hline
  & \multicolumn{3}{c|}{D-MORPH}&\multicolumn{3}{c|}{DE} \\
\hline
  $\alpha$ & $F_1$ & $F_2$ &$F$ & $F_1$ & $F_2$ & $F$\\
\hline
 0.05 &0.9856 & 0.7352 & 1.0224 & 0.9999 & 0.9093 &1.0451\\
  \hline
 0.1 & 0.9680 & 0.7433 & 1.0423 & 0.9997 & 0.9036 &1.0901\\
  \hline
 0.2 & 0.9460 & 0.7519 & 1.0964 & 0.9992 & 0.9071 &1.1806\\
  \hline
 0.4 & 0.9287 & 0.7534 & 1.2300  &0.9980 & 0.9132 &1.3633\\
  \hline
 0.6 & 0.9210 & 0.7541 & 1.3735  &0.9963 & 0.9161 &1.5460\\
  \hline
 0.8 & 0.9159 & 0.7555 & 1.5203  &0.9958 & 0.9150 &1.7278\\
  \hline
 1.0 & 0.9119 & 0.7568 & 1.6688  &0.9938 & 0.9193 &1.9131\\
  \hline
\end{tabular}
\caption{Optimization results searched by D-MORPH and DE for a Landau-Zener Hamiltonian. The target state fidelity ($F_1$) and the averaged system energy during the evolution ($F_2$) are shown with different weight factors $\alpha$. The adiabatic time was chosen as $T=3$.}\label{table1}
\end{table}

\begin{figure*}
  \centering
\includegraphics[width=0.8\textwidth,height=0.6\textwidth]{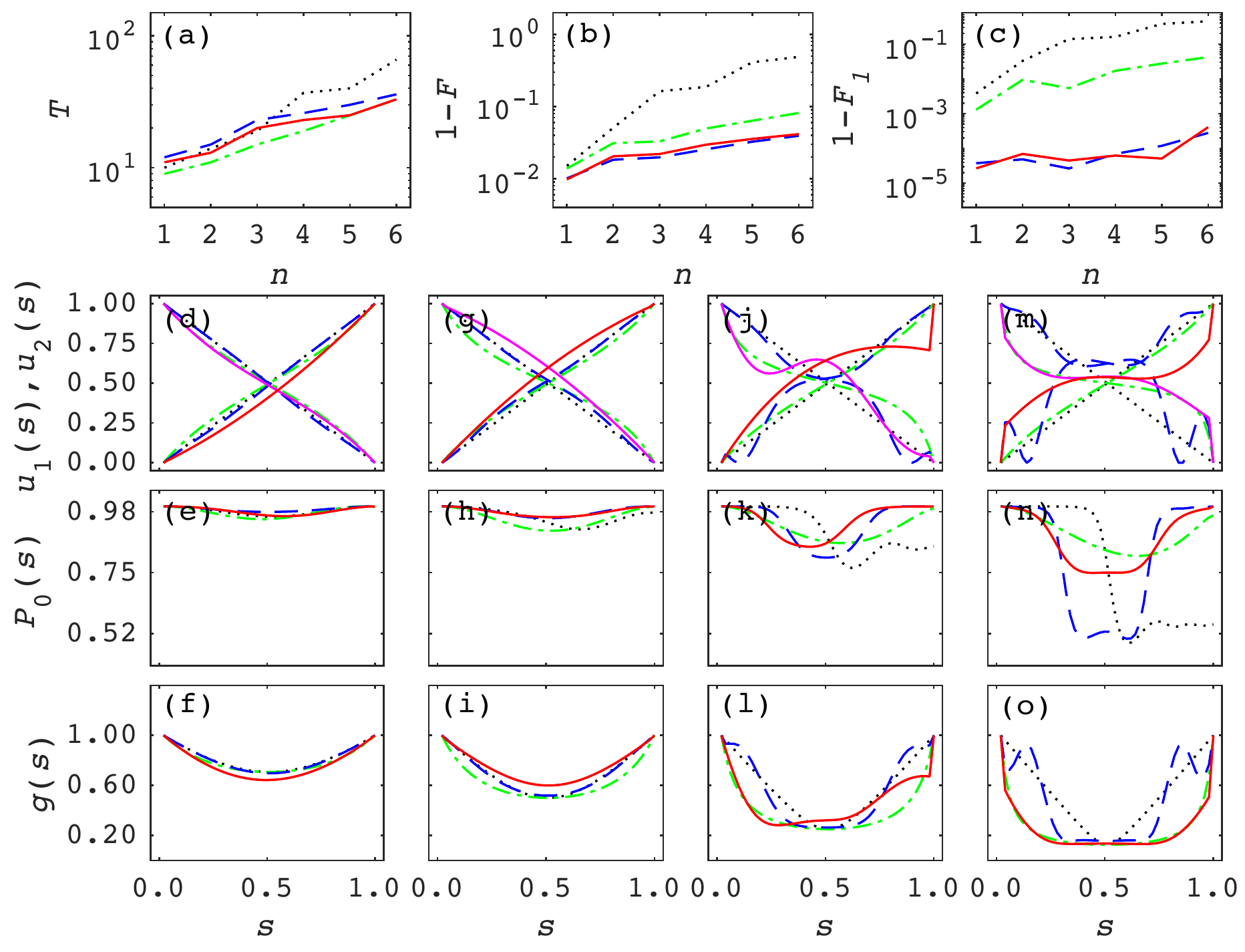}\\
  \caption{(Color online) Optimization results for the Grover search algorithm Hamiltonian using $n=1$ to $6$ qubits obtained by the Linear, RC, D-MORPH, and DE methods. (a) The searched minimum adiabatic time $T$ vs the qubit number $n$ when $\alpha=0.1$, where $T$ was gradually increased and stopped when the difference between two successive $F$ was smaller than $10^{-3}$. (b), (c) The corresponding searched final performance function values $1-F$ and $1-F_1$ vs $n$. The control fields $u_1(s)$ and $u_2(s)$, the instantaneous ground-state population $P_0(s)$, and the energy gap $g(s)$ are shown vs the scaled time $s$ for (d)--(f) $n=1$; (g)--(i) $n=2$; (j)--(l) $n=4$; and (m)--(o) $n=6$. In all figures, the different methods are Linear (dotted black line), RC (dash-dotted green line), D-MORPH (dashed blue line), and DE (solid red line), and all the algorithm parameters were the same as the above Laudau-Zener Hamiltonian case. }\label{Figure3}
  \end{figure*}

In the following, from the above simulations we chose two sets of the combinations of $T$ and $\alpha$ to demonstrate the controls fields, the instantaneous ground-state population and the energy gap obtained by the Linear, D-MORPH, and DE methods, as shown in Fig. \ref{Figure2}. By comparing the instantaneous ground-state population in Figs. \ref{Figure2}(b) and \ref{Figure2}(h) for D-MORPH and that in Figs. \ref{Figure2}(e) and \ref{Figure2}(k) for DE, we find that during the evolution, DE has a generally smaller population loss when $\alpha=0.1$ and $\alpha=0.5$, and both of them beat the Linear method. Moreover, from the comparison of Figs. \ref{Figure2}(c) and \ref{Figure2}(i) and Figs. \ref{Figure2}(f) and \ref{Figure2}(l), we find that the energy gap induced by DE is almost the inverse of that induced by Linear for both cases $\alpha=0.1$ and $0.5$, and it can be greater than $2$ at all times during the evolution when $\alpha=0.1$. However, the energy gap induced by D-MORPH is similar to that of Linear when $\alpha=0.1$ and $\alpha=0.5$. These results indicate that the improved performance is achieved by a gap increment at intermediate times.

\subsection{Grover Search Algorithm Hamiltonian}

We also considered a more practical and complex example, namely, the Grover search algorithm, which is used to identify a marked element in an unsorted database of $N$ elements. Precisely speaking, its Hamiltonian can be denoted as $\mathcal H_I=\mathcal{I}-| \varphi \rangle \langle \varphi |, \mathcal H_P=\mathcal{I}-| m \rangle \langle m |$, where $\mathcal{I}$ is the identity matrix, $|\varphi \rangle$ is the uniform superposition state $|\varphi \rangle=\sum_{i = 0}^{N - 1} {\left| i \right\rangle}/{\sqrt N} $, $\{| i \rangle\}$ are the basis states of the Hilbert space, and $|m \rangle$ is the marked state. Local adiabatic evolution theorem \cite{RC02} based RC promises an adiabatic time of the order of $\sqrt{N}$, which is a quadratic speed-up compared to the classical Linear method. The optimization algorithms D-MORPH and DE are expected to surpass or at least be close to this scaling. 
 
 Thus, we first explored the minimum adiabatic time $T$ needed to reach a sufficiently high $F$ versus the number of qubits $n$ for these methods; the results are shown in Fig. \ref{Figure3}(a). One can find that D-MORPH, DE, and RC all have a quadratic speed-up compared to Linear, as expected. Moreover, the adiabatic time needed for DE is always smaller than that for D-MORPH, which indicates that DE can achieve a faster adiabatic evolution than D-MORPH. To show the corresponding performance function values $F$ and $F_1$ obtained by these methods, we plot Figs. \ref{Figure3}(b) and \ref{Figure3}(c), from which we find that DE achieves a comparable multi objective function value $F$ with D-MORPH. For the state fidelity $F_1$, DE also has a comparable performance with D-MORPH for most of the cases.  
 
We then proceed by demonstrating the control fields, the instantaneous ground-state population, and the energy gap obtained by these methods for the number of qubits $n=1,2,4,6$, as shown in the rest of Fig. \ref{Figure3}. The instantaneous ground-state population comparisons plotted in Figs. \ref{Figure3}(e),\ref{Figure3}(h),\ref{Figure3}(k), and \ref{Figure3}(n) reveal that DE performs much better for reducing the population loss during the evolution compared to D-MORPH, especially for a large number of qubits, i.e., $n=4,6$. The corresponding energy-gap comparisons shown in Figs. \ref{Figure3}(f),\ref{Figure3}(i),\ref{Figure3}(l), and \ref{Figure3}(o) report generally similar behaviors of all the methods, suggesting that we may need more careful research on adjusting the energy gaps by the searched optimal control schedules to further improve the adiabatic quantum pathways. By exploring the adiabatic pathways of the Grover search algorithm Hamiltonian by Linear, RC, D-MORPH, and DE, we can conclude that DE achieves almost the fastest adiabatic evolution, while achieving the least instantaneous ground-state population loss. 

\begin{figure}
  \centering
\includegraphics[width=0.48\textwidth,height=0.28\textwidth]{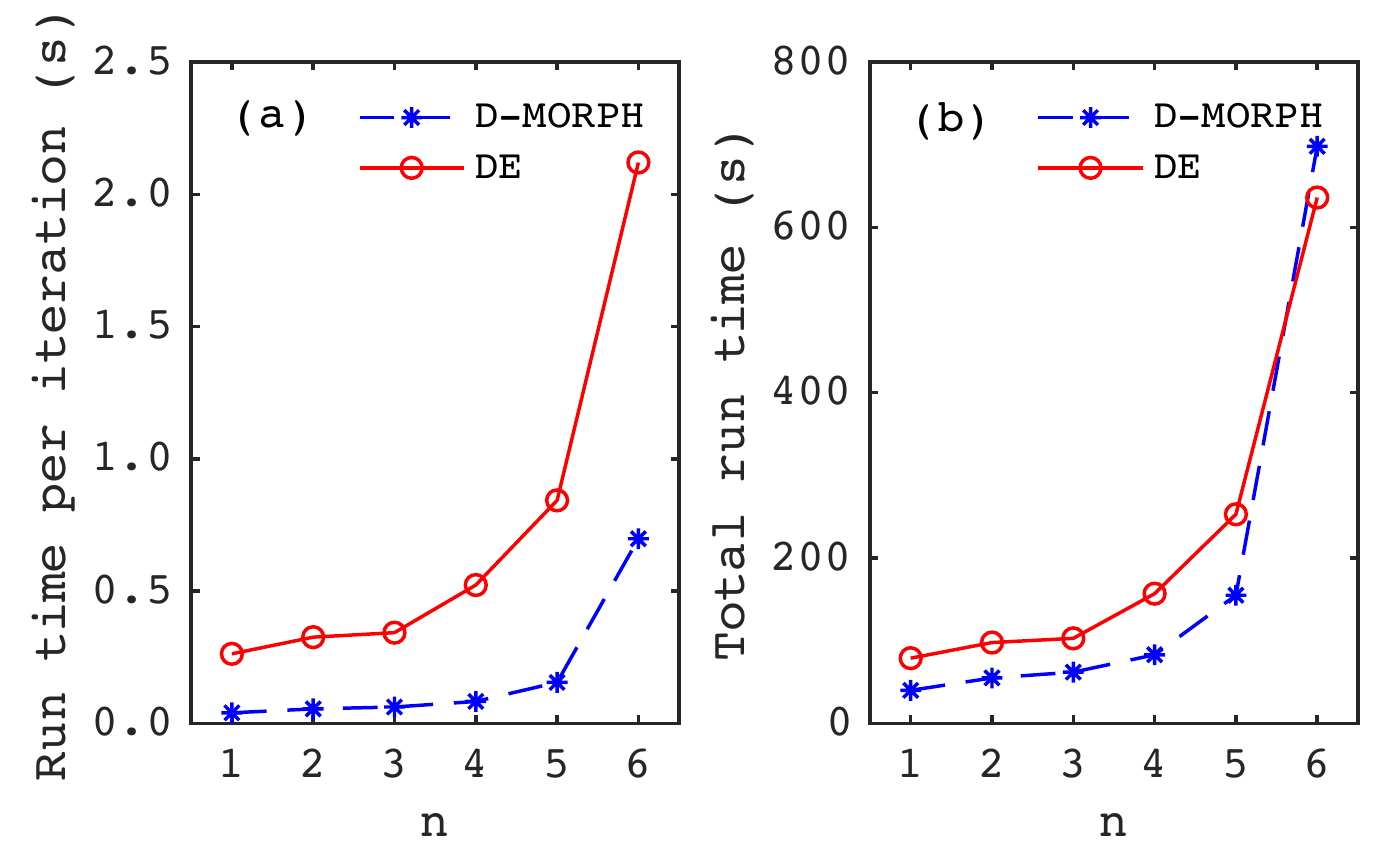}\\
  \caption{(Color online) Run time of D-MORPH and DE for the Grover search algorithm Hamiltonian. (a) The run time per iteration with respect to the number of qubits $n$. (b) The total run time regarding the number of qubits $n$.}\label{Figure4}
  \end{figure}

In addition, we briefly analyze the computational costs of D-MORPH and DE in searching optimal control schedules here. They are partly determined by the algorithm parameters, including $S,C,P,D,N_c$ for DE and $\lambda^G$ for D-MORPH, which are very important for the performance of the algorithms. However, a thorough tuning of the parameters will be a resource-consuming and unrealistic task. For our simulations, $S,C,P$ are chosen from our experiences and $D,N_c,\lambda^G,G_{\max}$ are settled by sufficient trials. In this way, we expect that D-MORPH and DE perform possibly close to their best status, respectively. We then show the run time per iteration and the total run time for these two methods in Fig. \ref{Figure4}, where we find that DE needs significantly more run time per iteration than that for D-MORPH. However, the total run time for DE is a little longer than that for D-MORPH when the number of qubits $n=1 \sim 5$, and shorter when $n=6$. From this analysis, we can roughly conclude that the computational costs of the two methods are comparable and both are in an acceptable range.

\section{Conclusions and discussions} \label{Conclusion}
In this study, we have proposed a differential evolution algorithm with the CRAB technique to explore the optimal adiabatic quantum pathways for AQC and apply it to a Landau-Zener Hamiltonian and Grover search algorithm Hamiltonian. This gradient-free learning algorithm performs better than conventional methods, including Linear and RC, that are based on adiabatic theorems. This is because  most of the adiabatic theorems are not exact so that their induced adiabatic pathways are approximate. Even these conventional approaches can give nearly optimal solutions, and an easier-to-implement numerical method will be more friendly to applications. Moreover, compared to a recent gradient-based D-MORPH method, our method also has advantages in terms of realizing  the high-fidelity target ground state with shorter adiabatic time and reducing the population loss from the instantaneous ground state. The merits of our gradient-free method mainly come from two reasons \cite{PS06}: (1) For multiobjective optimization problems, the landscape of the performance function usually contains many local extrema. Gradient-based algorithms start from one trial point and move along the derivative direction, thus are very likely to get trapped in these local extrema. However, evolutionary-based algorithms start from a group of points distributed in the whole parameter space and update according to some evolutionary rules, thus having more chance to escape from these local extrema and reach the global optima. (2) The to-be-optimized adiabatic quantum pathways contain amplitude constraints (in the range $[0,1]$), which greatly influences the performance of the optimization algorithms and also induces local extrema \cite{MKW12}. For the gradient-based type, the amplitudes of the control fields vary depending on the continuous derivative functions, and thus the amplitude constraints will very likely induce the convergence to the false extreme. However, for gradient-free algorithms, the amplitudes change with more degrees of freedom, and they can be finely tuned to reach the true global optima. 

This numerical optimization method can handle multiobjective problems and constraints more easily; the complexity of the searching procedures for the optimal pathways does not increase much for more complex problems, and thus is more practical and useful for applied AQC. The successful applications here encourage us to extend it to more complicated AQC optimization tasks, such as satisfiability problems \cite{FG01}; and random optimization problems \cite{BF13}. Moreover, the proposed method can become an important tool for developing current quantum annealing hardware and future AQC processors, such as D-wave systems \cite{JAG11}.

Moreover, in real applications, analytical or numerically designed adiabatically quantum pathways may not behave as expected due to inevitable perturbations. However, our method can be easily adapted to the closed-loop type to handle these perturbations. This is because the performance function chosen here can be efficiently measured and the learning algorithm is resource saving compared to those gradient-based types.

\section*{ACKNOWLEDGEMENTS}
X.P. is supported by National Key Research and Development Program of China (Grant No. 2018YFA0306600), the National Science Fund for Distinguished Young Scholars (Grant No. 11425523), Projects of International Cooperation and Exchanges NSFC (Grant No. 11661161018), and Anhui Initiative in Quantum Information Technologies (Grant No. AHY050000). J.L. is supported by the National Natural Science Foundation of China (Grants No. 11975117, No. 11605005, No. 11875159, and No. U1801661), Science, Technology and Innovation Commission of Shenzhen Municipality (Grants No. ZDSYS20170303165926217, No. JCYJ20170412152620376, and No. JCYJ20180302174036418)), and Guangdong Innovative and Entrepreneurial Research Team Program (Grant No. 2016ZT06D348).


\providecommand{\noopsort}[1]{}\providecommand{\singleletter}[1]{#1}%

\end{document}